\documentclass[twocolumn,aps,pre,floatfix,superscriptaddress,longbibliography]{revtex4-1}
\pdfoutput=1 %This is for the ArXiv submission only
\usepackage{amsmath,amssymb,eucal}
\usepackage{graphicx}

\usepackage[colorlinks=true, urlcolor=blue, anchorcolor=blue, citecolor=blue,filecolor=blue,linkcolor=blue,menucolor=blue,pagecolor=blue]{hyperref}

\begin{document}
\title{Aggregation-Fragmentation Processes with Broken Detailed Balance}

\author{P.~L.~Krapivsky}
\affiliation{Department of Physics, Boston University, Boston, Massachusetts 02215, USA}
\affiliation{Santa Fe Institute, Santa Fe, New Mexico 87501, USA}

\begin{abstract} 
We study aggregation-fragmentation processes in which pairs of clusters can aggregate, and each cluster can break into two fragments. If the rates of aggregation and fragmentation do not depend on the masses, detailed balance does not hold, but nonequilibrium steady states can still be deduced from an exact solution for the Laplace transform. For models in which aggregation rates remain constant but fragmentation rates scale as $(\text{mass})^\beta$, detailed balance holds only when $\beta=1$. Away from this solvable case, we employ asymptotic techniques and show that when $\beta\geq 0$, the steady states share similarities with those from the mass-independent ($\beta=0$) model. An instantaneous shattering transition with continuous mass loss occurs when $\beta<0$. 
\end{abstract}

\maketitle

\section{Introduction}
\label{sec:Intro}

In pure aggregation, clusters merge irreversibly, leading to a cascade toward larger sizes. Aggregation and fragmentation often act together, and the competition between these two processes usually leads to nonequilibrium steady states \cite{BT45,Melzak57,AB79,strings,Vigil,book}. (Exotic behaviors with never-ending oscillations have been also observed \cite{Sergey17,Sergey18,Pego20,burst,Stas21,Fagan21,Pego22,Fortin,Fagan24}.) It is easy to determine nonequilibrium steady states when the detailed balance condition holds \cite{Vigil,book}. The detailed balance is valid in isolated models, yet the lack of detailed balance is a generic phenomenon \cite{book,BK08}.  The detailed balance is violated in the chipping model \cite{KR96,Satya98,IK, Rajesh01,Jain,Rajesh02} and generally in the models where some reaction rates vanish. However, one could hope that detailed balance holds in models with natural reaction rates akin to the models studied in \cite{BT45,Melzak57,AB79,strings, Vigil,book}. Below, we show that the detailed balance does not hold even in models more natural than the model solved by Blatz and Tobolsky \cite{BT45} in the first study of aggregation-fragmentation processes. Specifically, for aggregation-fragmentation processes with mass-independent rates, the stationary solutions do not follow from the detailed balance.

For all processes analyzed below, we postulate that aggregation proceeds with mass-independent rates. Without fragmentation, the model reduces to the classical aggregation process. This process was analyzed by Smoluchowski \cite{Smol17} in the first study of aggregation and remains a rare analytically solvable aggregation process \cite{book,Chandrasekhar43,Flory,Drake,Leyvraz03}. When this aggregation process is supplemented by binary fragmentation occurring at a rate proportional to the cluster mass, the detailed balance condition is satisfied. Using it, one derives the nonequilibrium steady state \cite{BT45}. Here, we investigate the models with mass-independent fragmentation rates. One might expect that these models are more tractable than the classical aggregation-fragmentation processes \cite{BT45}, but this is not the case. The reason is the violation of the detailed balance. Therefore, the steady state is complicated, and elucidating its properties requires substantial effort.   

Mass-independent splitting can be either random or deterministic. In the context of animal-group statistics, particularly the stationary distributions of schools of fish \cite{Okubo86,Okubo-Levin,Niwa96,Bonabeau,Niwa04,Ma11,Pego}, the splitting was random. These studies employed a continuous-time dynamics; both discrete and continuous mass distributions have been considered. The trail dynamics \cite{Trail:1d} proceeds on the one-dimensional lattice. Each site contains at most one cluster, and clusters hop symmetrically to nearest-neighbor sites with mass-independent hopping rates; if a cluster hops to an already occupied site, both clusters instantaneously merge. In the case of random splitting, a cluster of mass $x$ splits, $x\to \{ax,(1-a)x\}$, with splitting parameter $a$ chosen randomly and uniformly from the interval $(0,1)$; splitting parameters in different splitting events are uncorrelated. In the deterministic splitting, $x\to \{rx,(1-r)x\}$, the number $r\in (0,1)$ is fixed; for instance, if $r=1/2$, a cluster always splits into two equal halves:  $x\to \{x/2,x/2\}$. A discrete-time dynamics employed in Ref.~\cite{Trail:1d} depends on $r\in (0,1)$ and $p\in (0,1)$ defined as the probability of the splitting event;  the hopping probability was taken to be $1-p$.  (Reference \cite{Trail24} studies a mathematically similar model describing flux-conserving directed percolation.) 

We employ a continuous-time formulation that is physically more natural than a discrete-time framework. We treat mass as a continuous variable. This choice is customary in studies of pure fragmentation \cite{Filippov,MZ85,MZ86,MZ87,Redner90,Ernst93} and often appropriate in modeling aggregation-fragmentation processes. In some applications, however, clusters consist of an integer number of elemental building blocks, so mass is a discrete variable. In problems involving fragmentation, the results in the continuous and discrete mass settings are usually qualitatively similar; in the discrete mass setting, the results tend to be more cumbersome than in the continuum setting. We also employ a mean-field description, i.e., we tacitly assume that the system remains well-mixed. The validity of the mean-field framework for describing aggregation in low spatial dimensions is questionable \cite{book}. However, if the steady state emerges, the major discrepancy between the mean-field and exact predictions is in the approach to the steady state that is inaccurately accounted for by the mean-field framework. 

The model with random splitting is more tractable. In Sec.~\ref{sec:US}, we present its complete solution. For the model with deterministic splitting (Sec.~\ref{sec:DS}), the qualitative behavior of the steady-state mass distribution is the same as in the model with random splitting. We also derive major quantitative features of the steady-state mass distribution in the small- and large-mass limits. Completing the analytical description requires solving neat functional equations. In the simplest case of splitting into equal halves, we reduce these functional equations to logistic maps with some growth rates. Despite 150 years of effort, explicit analytical solutions to the logistic map were found only for three special values of the growth rate, different from the ones appearing in our problem. 

We then consider a class of models with homogeneous random splitting: The overall splitting rate scales algebraically with cluster mass, namely as $(\text{mass})^\beta$. The models with $\beta\geq 0$ analyzed in Sec.~\ref{sec:hom} exhibit qualitatively similar behaviors to the model with $\beta=0$ analyzed in Sec.~\ref{sec:US}. The detailed balance is obeyed only for the model with $\beta=1$, the first studied aggregation-fragmentation process \cite{BT45}. We employ the Laplace transform and asymptotic analysis to extract quantitative properties of the nonequilibrium steady states in models with $\beta>0$. 

When $\beta < 0$, mass is not conserved because the cascading breakup rate creates increasingly smaller fragments. These fragments ultimately result in dust---an infinite number of zero-mass fragments representing a finite fraction of the mass. This shattering transition is instantaneous as it begins at $t=+0$ in our models (Sec.~\ref{sec:minus}). Pure fragmentation processes with homogeneous splitting rates also undergo an instantaneous shattering transition when $\beta < 0$. These fragmentation processes are analytically tractable (Appendix~\ref{ap:fragm}) as the governing equations are linear. The nonlinear integro-differential equations for aggregation-fragmentation processes are not solvable and even appear mathematically inconsistent when $-1\leq \beta<0$, as the cluster density appearing in these equations is infinite. When $\beta<-1$, the governing equations are well-defined. We analyze these equations by assuming that solutions take a scaling form in the long-time limit. While results are consistent, they are incomplete. Key asymptotic predictions include $t^{-1}$ decay of cluster density and $t^{-1-1/\beta}$ decay of mass density. The universal (independent of $\beta$) cluster density decay differs from the less universal behaviors in pure fragmentation processes. Sec.~\ref{sec:disc} summarizes our findings and outlines future research avenues.

\section{Random Splitting}
\label{sec:US}

We work in the continuum setting, i.e., we treat mass as a real number. We postulate that mass fully characterizes each cluster, i.e., we ignore other characteristics, such as cluster shape. The density $c(x,t)$ of clusters of mass $x$ at time $t$ satisfies an integro-differential equation
\begin{eqnarray}
\label{AF}
\frac{\partial c(x,t)}{\partial t} &=& \int_0^x dy\,c(y,t)c(x-y,t) - 2N(t) c(x,t) \nonumber\\
                                             & + & 2\lambda \int_x^\infty dz\,\frac{c(z,t)}{z} - \lambda c(x,t)
\end{eqnarray}
where
\begin{equation}
N(t) = \int_0^\infty dz\,c(z,t)
\end{equation}
is  the total cluster density. The first two terms on the right-hand side of \eqref{AF} describe aggregation events; we set the mass-independent merging rate to unity. The last two terms account for fragmentation events. The total splitting rate is also mass-independent; we denote it by $\lambda$. (With our choice of the merging rate, $\lambda$ is the ratio of the total splitting rate to the merging rate.) 

Equations \eqref{AF} and equations describing similar aggregation-fragmentation process with discrete masses have been studied in \cite{Pego}. The goal was to describe animal group-size statistics \cite{Okubo86,Okubo-Levin}, particularly the size distribution of schools of fish \cite{Niwa96,Bonabeau,Niwa04,Ma11}. 

Our analysis is more concise, and we present it to explain our approach in the realm of the simplest model. We demonstrate the usefulness of the moments of the steady-state mass distribution and the power of the Laplace transform in the problem. The same techniques apply to the model with deterministic splitting (Sec.~\ref{sec:DS}) and to a class of models with random splitting and the splitting rate algebraically varying with mass (Sec.~\ref{sec:hom}). 

Mass is conserved in merging and splitting events. We set the mass density to unity: 
\begin{equation}
\label{mass}
M(t) = \int_0^\infty dz\,z c(z,t) = 1
\end{equation}
This can always be done by rescaling: $c\to c/M$.

Integrating \eqref{AF} over $0<x<\infty$ we find a closed rate equation for the cluster density: 
\begin{equation}
\label{NNNN}
\frac{dN}{dt}=-N^2+\lambda N
\end{equation}
 Solving \eqref{NNNN} yields
\begin{equation}
\label{N:sol}
N(t) = \frac{\lambda}{1+\left[\frac{\lambda}{N(0)}-1\right]e^{-\lambda t}}
\end{equation}
showing that the cluster density quickly approaches to
\begin{equation}
\label{NL}
N(\infty) = \lambda
\end{equation}

Multiplying Eq.~\eqref{AF} by $x^2$ and integrating we deduce a simple rate equation
\begin{subequations}
\begin{equation}
\label{M2}
\frac{dM_2}{dt}= 2 - \frac{\lambda}{3}\,M_2
\end{equation}
for the second moment $M_2(t)=\int_0^\infty dx\,x^2 c(x,t)$. Equation \eqref{M2} is also closed, and solving it gives 
\begin{equation}
\label{M2:sol}
M_2(t) = \frac{6}{\lambda} +\left[M_2(0)-\frac{6}{\lambda}\right]e^{-\lambda t/3}
\end{equation}
\end{subequations}
Similarly one finds
\begin{subequations}
\begin{equation}
\label{M3}
\frac{dM_3}{dt}= 6 M_2 - \frac{\lambda}{2}\,M_3
\end{equation}
for the third moment $M_3(t)=\int_0^\infty dx\,x^3 c(x,t)$. Thus 
\begin{eqnarray}
\label{M3:sol}
M_3(t) &=& \frac{72}{\lambda^2} +\frac{36}{\lambda}\left[M_2(0)-\frac{6}{\lambda}\right]e^{-\lambda t/3} \nonumber\\
            &+&\left[M_3(0)-\frac{36}{\lambda}\,M_2(0)+ \frac{144}{\lambda^2}\right]e^{-\lambda t/2}
\end{eqnarray}
\end{subequations}
Therefore, the second and third moments of the mass distribution also quickly approach the stationary values: $M_2(\infty) = 6/\lambda$ and $M_3(\infty) = 72/\lambda^2$. 

The above examples suggest that the mass distribution becomes stationary as $t\to\infty$. Equating the time derivative in \eqref{AF} to zero and using \eqref{NL} we find that the stationary density $c(x)\equiv c(x,\infty)$ satisfies 
\begin{equation}
\label{AF:uniform}
3 c = \lambda^{-1} c*c + 2\int_x^\infty dz\,\frac{c(z)}{z}
\end{equation}
where we have used the notation  
\begin{equation}
\label{conv}
c*c= \int_0^x dy\,c(y)c(x-y) 
\end{equation}
for the convolution. The appearance of the convolution in Eq.~\eqref{AF:uniform} suggests to perform the Laplace transform of the stationary mass distribution
\begin{equation}
\label{LT:def}
C(s) = \int_0^\infty dx\,e^{-sx} c(x)
\end{equation}
 We find that $C(s)$ is the root of the cubic polynomial 
\begin{equation}
\label{solution}
s[C(s)]^3  = \lambda^3[\lambda-C(s)]
\end{equation}
from which we deduce the asymptotic behaviors 
\begin{subequations}
\label{small-large}
\begin{align}
\label{small}
&c(x)\simeq \frac{\lambda^{4/3}}{\Gamma\big(\frac{1}{3}\big)}\,x^{-2/3} \qquad  \qquad  \qquad  ~~\quad  x\ll \lambda^{-1}\\
\label{large}
&c(x)\simeq \frac{9}{8}\,\sqrt{\frac{\lambda}{\pi}}\, x^{-3/2}\, \exp\!\left[-\frac{4}{27}\,\lambda x\right] \quad x\gg \lambda^{-1}
\end{align}
\end{subequations}
The details of the derivation of \eqref{solution}--\eqref{small-large} are presented in Sec.~\ref{subsec:Lap}. 

The dependence on $\lambda$ can be understood using scaling arguments alone. Indeed, the transformation
\begin{equation}
\label{transform}
x=X/\lambda, \quad c(x) = \lambda^2 F(X)
\end{equation}
recasts the governing equation \eqref{AF:uniform} into
\begin{equation}
\label{AF1:uniform}
3 F =  F*F+ 2\int_X^\infty dZ\,\frac{F(Z)}{Z}
\end{equation}
The mass conservation, Eq.~\eqref{mass}, becomes 
\begin{equation*}
 \int_0^\infty dX\,XF(X) = 1
\end{equation*}
The asymptotic behaviors \eqref{small}--\eqref{large} of $c(x)$ lead to independent on $\lambda$ asymptotic behaviors of $F(X)$ 
\begin{equation}
\label{FX:asymp}
F(X) \simeq
\begin{cases}
1/\big[\Gamma(1/3)\,X^{2/3}\big]                           &X\ll 1\\
 \frac{9}{8\sqrt{\pi}}\,X^{-3/2}\, e^{-4X/27}   &X\gg 1
\end{cases}
\end{equation}

Before deriving the asymptotic behaviors \eqref{small}--\eqref{large}, equivalently \eqref{FX:asymp}, we deduce the exact stationary values of the moments.  

\subsection{Moments}

We have already established the steady-state moments of the mass distribution 
\begin{equation}
\label{Mn}
M_n = \int_0^\infty dz\,z^n c(z) 
\end{equation}
for small $n$: $M_0=N=\lambda, M_1=M=1, M_2=6/\lambda$ and $M_3 = 72/\lambda^2$. To determine higher moments we multiply \eqref{AF:uniform} by $x^n$ and integrate over $0<x<\infty$ to yield 
\begin{equation}
\label{moments}
\sum_{m=1}^{n-1} \binom{n}{m} M_mM_{n-m} = \lambda\,\frac{n-1}{n+1}\, M_n
\end{equation}
Specializing \eqref{moments} to $n=2$ we confirm $M_2=6/\lambda$ which is then used to recover $M_3=72/\lambda^2$. Generally
\begin{equation}
\label{MA}
M_n = \frac{A_n}{\lambda^{n-1}}
\end{equation}
The amplitudes can be found from the recurrence
\begin{equation}
\label{Amn}
\sum_{m=1}^{n-1} \binom{n}{m} A_m A_{n-m} = \frac{n-1}{n+1}\, A_n
\end{equation}
Here are the amplitudes $A_n$ with $n\leq 11$:
\begin{eqnarray*}
&& A_0 = A_1 = 1, ~~A_2 = 6, ~~ A_3 = 72, ~~ A_4 = 1\,320 \\
&& A_5 = 32\,760, ~~~A_6 = 1\,028\,160, ~~ A_7 = 39\,070\,080\\
&& A_8 = 1\,744\,364\,160, ~~A_9 = 89\,513\,424\,000\\
&&  A_{10} = 5\,191\,778\,592\,000, ~~A_{11}=335885501952000
\end{eqnarray*}
The On-Line Encyclopedia of Integer Sequences suggests that these elements belong to the sequence 
\begin{equation}
\label{An}
A_n = \frac{(3n)!}{(2n+1)!}
\end{equation}
and indicates \footnote{This is sequence A001763 in the On-line Encyclopedia of Integer Sequences; it represents the number of dissections of a ball.} an interesting geometric interpretation of the sequence $A_n$. To verify the guess \eqref{An} we introduce a generating function
\begin{equation}
\label{Az:def}
\mathcal{A}(z) = \sum_{n\geq 1} \frac{A_n z^n}{n!}
\end{equation}
and recast the recurrence \eqref{Amn} into an integral equation 
\begin{equation}
\label{Az:eq}
\mathcal{A}^2(z) = \mathcal{A}(z)  - \frac{2}{z}\int_0^z dw\, \mathcal{A}(w)
\end{equation}
The generating function \eqref{Az:def} simplifies to 
\begin{equation}
\label{Az:sol}
\mathcal{A}(z) = -1 + \frac{2}{\sqrt{3z}}\,\sin\!\left[\frac{1}{3}\,\sin^{-1}\!\left(\sqrt{\frac{27z}{4}}\right)\right]
\end{equation}
if the sequence $A_n$ is given by \eqref{An}. A lengthy calculation shows that \eqref{Az:sol} is the solution of Eq.~\eqref{Az:eq}. 

The expression \eqref{Az:sol} holds on the interval $0<z<\frac{4}{27}$ and exhibits a singular behavior 
\begin{equation}
\label{Az:sing}
\mathcal{A}(z) = \frac{1}{2}-\frac{\sqrt{3}}{2}\,\sqrt{1-\frac{27z}{4}}+\ldots
\end{equation}
in the $z\uparrow \frac{4}{27}$ limit. The singularity reflects the large $n$ behavior of the sequence $A_n$:
\begin{equation}
\label{An:asymp}
A_n \simeq \sqrt{\frac{3}{16\pi}}\,\frac{n!}{n^{3/2}}\left(\frac{27}{4}\right)^n
\end{equation}

\subsection{Small Mass Tail}

In the $X\to 0$ limit one can drop the convolution term in \eqref{AF1:uniform}. We thus obtain an integral equation 
\begin{equation}
3F(X) \simeq 2\int_X^\infty dZ\,\frac{F(Z)}{Z}
\end{equation}
which we differentiate to yield $3F'=-2F/X$ leading to 
\begin{equation}
\label{small:A}
F(X) \simeq A X^{-2/3} \qquad\text{when}\quad X\to 0
\end{equation}
This elementary approach does not give the amplitude $A$. To fix $A=1/\Gamma\big(\frac{1}{3}\big)$ we derive the exact Laplace transform. 

\subsection{Laplace Transform}
\label{subsec:Lap}

Performing the Laplace transform of Eq.~\eqref{AF:uniform} yields
\begin{equation}
\label{LT}
3 C = \frac{C^2}{\lambda}+2 \int_0^\infty dz\,c(z)\,\frac{1-e^{-sz}}{sz}
\end{equation}
with $C$ defined by \eqref{LT:def}. Multiplying \eqref{LT} by $s$ and then differentiating with respect to $s$ we get
\begin{equation*}
C\,\frac{\lambda - C}{3\lambda - 2C}+s\,\frac{d C}{ds} = 0
\end{equation*}
which we integrate to find
\begin{equation}
\label{sol}
\ln s + 3\ln C -\ln[\lambda-C] = \text{const}
\end{equation}
Expanding \eqref{LT:def} at $s\to 0$ we obtain
\begin{equation}
C(s)=\sum_{n\geq 0}M_n \,\frac{(-s)^n}{n!}=\lambda-s+O(s^2)
\end{equation}
Substituting this expansion into \eqref{sol}, we find that the constant on the right-hand side is $3\ln\lambda$. Thus Eq.~\eqref{sol} becomes
\begin{equation}
\ln s + 3\ln C -\ln[\lambda-C] = 3\ln\lambda
\end{equation}
leading to the announced result \eqref{solution} for the Laplace transform.

The small-mass behavior of $c(x)$ can be extracted from the asymptotic behavior of the Laplace transform in the $s\to\infty$ limit. Using \eqref{solution} we deduce $C\simeq \lambda^{4/3} s^{-1/3}$, while \eqref{small:A} implies $C\simeq A\lambda^{4/3}\Gamma\big(\frac{1}{3}\big) s^{-1/3}$. Thus the amplitude in \eqref{small:A} is $A=1/\Gamma\big(\frac{1}{3}\big)$ giving the announced asymptotic \eqref{small}.

The Laplace transform $C(s)$ is a root of the cubic polynomial \eqref{solution}. For $s>0$, the proper root is
\begin{subequations}
\begin{align}
\label{Cs}
C(s) &=\left(\frac{\lambda^4}{2s}\right)^{1/3}F(s)- \left(\frac{2\lambda^5}{27 s^2}\right)^{1/3}\frac{1}{F(s)}\\
\label{F+}
F(s) &= \left\{1+\sqrt{1+\frac{4}{27}\,\frac{\lambda}{s}}\right\}^{1/3}
\end{align}
\end{subequations}
One can use \eqref{Cs}--\eqref{F+} to re-derive \eqref{small}, but the above direct calculation relying on \eqref{solution} seems easier. 

\subsection{Large Mass Tail}

Equation \eqref{F+} hints that the Laplace transform remains well-defined only for $s > -\frac{4}{27}\,\lambda$. [The divergence of the moment generating function \eqref{Az:sol} when $z>\frac{4}{27}$ has the same cause.] Instead of using a cumbersome expression for the root, we employ a perturbation treatment. We notice that at $s_* = -\frac{4}{27}\,\lambda$, the proper root of \eqref{solution} is $C(s_*)=3\lambda/2$. We then write
\begin{equation}
s=s_*(1+\epsilon)^{-1}, \quad C(s)=\frac{3\lambda}{2}\,(1+\delta)
\end{equation}
Inserting this into \eqref{solution} and treating $\epsilon$ and $\delta$ as small we obtain $\delta\simeq -\sqrt{\epsilon/3}$ and hence 
\begin{equation}
\label{C:sing}
C(s) - \frac{3\lambda}{2} \simeq -\frac{9}{4}\,\sqrt{\lambda}\,\sqrt{s-s_*}
\end{equation}
in the leading order. The derivative diverges: 
\begin{equation}
\label{Cs:der}
\frac{dC}{ds}  \simeq -\frac{9}{8}\,\frac{\sqrt{\lambda}}{\sqrt{s-s_*}}
\end{equation}
as $s\to s_*+0$. This asymptotic behavior of the derivative of the Laplace transform implies that the large-mass tail is exponential with an algebraic pre-factor:
\begin{equation}
\label{large:B}
c(x)\simeq B x^{-3/2}\, e^{s_* x}
\end{equation}
Indeed, \eqref{large:B} implies that in the $s\to s_*+0$
\begin{equation}
\label{Cs:der2}
\frac{d C}{ds}  \simeq -B\int_0^\infty \frac{dx}{\sqrt{x}}\,e^{-(s-s_*)x}
                                 = -\frac{B\sqrt{\pi}}{\sqrt{s-s_*}}
\end{equation}
Comparing \eqref{Cs:der} and \eqref{Cs:der2} we fix the amplitude $B=\frac{9}{8}\sqrt{\frac{\lambda}{\pi}}$ which in conjunction with \eqref{large:B} lead to the announced asymptotic behavior \eqref{large}.

\section{Deterministic Splitting}
\label{sec:DS}

In this section we consider the deterministic splitting 
\begin{equation}
x \to [rx, (1-r)x]
\end{equation}
where $r$ is a fixed number $r\in (0,1)$. 

The stationary mass distribution satisfies 
\begin{eqnarray}
\label{AF:r}
0 &=& \int_0^x dy\,c(y)c(x-y) - 2c(x) \int_0^\infty dz\,c(z) \nonumber\\
&+& \lambda\left[\frac{1}{r}\,c\Big(\frac{x}{r}\Big) + \frac{1}{1-r}\,c\Big(\frac{x}{1-r}\Big)  - c(x)\right]
\end{eqnarray}
The cluster density is again given by \eqref{NL} and hence \eqref{AF:r} simplifies to
\begin{eqnarray}
\label{AF:discrete}
3 c = \lambda^{-1}c*c+ \frac{1}{r}\,c\Big(\frac{x}{r}\Big) + \frac{1}{1-r}\,c\Big(\frac{x}{1-r}\Big)
\end{eqnarray}

The recurrence for the moments reads
\begin{equation}
\label{moments:r}
\frac{1}{\lambda}\sum_{m=1}^{n-1} \binom{n}{m} M_mM_{n-m} = \left[1-r^n-(1-r)^n\right]M_n
\end{equation}
The dependence on $\lambda$ is the same as before, Eq.~\eqref{MA}, so the amplitudes are determined from recurrence 
\begin{equation}
\label{Amn:r}
\sum_{m=1}^{n-1} \binom{n}{m} A_m A_{n-m} = \left[1-r^n-(1-r)^n\right]A_n
\end{equation}
starting with the ``initial condition'' $A_1=1$. The mirror symmetry, $r\leftrightarrow 1-r$, suggests using a manifestly symmetric splitting parameter $R=r(1-r)$ instead of $r$. The amplitudes have indeed significantly more compact expressions as functions of $R$ rather than $r$:
\begin{equation}
\label{A2-6}
\begin{split}
A_2 & = \frac{1}{R}\\ 
A_3 & = \frac{2}{R^2}\\ 
A_4 & = \frac{11}{R^3(2-R)}\\ 
A_5 & = \frac{2}{R^4(2-R)}\,\,\frac{19-4R}{1-R}\\ 
A_6 & = \frac{2}{7R^5(2-R)}\,\,\frac{473-333R+40R^2}{(1-R)^3}
\end{split} 
\end{equation}
etc. 

We now probe the tails of the stationary mass distribution. The small-mass tail is found from \eqref{AF:discrete} by dropping the convolution term,
\begin{equation}
\label{cr:small}
3 c(x) \simeq \frac{1}{r}\,c\Big(\frac{x}{r}\Big) + \frac{1}{1-r}\,c\Big(\frac{x}{1-r}\Big),
\end{equation}
and seeking the solution in the form
\begin{equation}
\label{cAr}
c(x)\simeq A(r) \lambda^{2-\rho} x^{-\rho}
\end{equation}
The dependence on $\lambda$ is again obvious as the transformation \eqref{transform} turns \eqref{AF:discrete} into equation independent on $\lambda$. The amplitude $A(r)$ depends only on the parameter $r$.

\begin{figure}
\centering
\includegraphics[width=8.1cm]{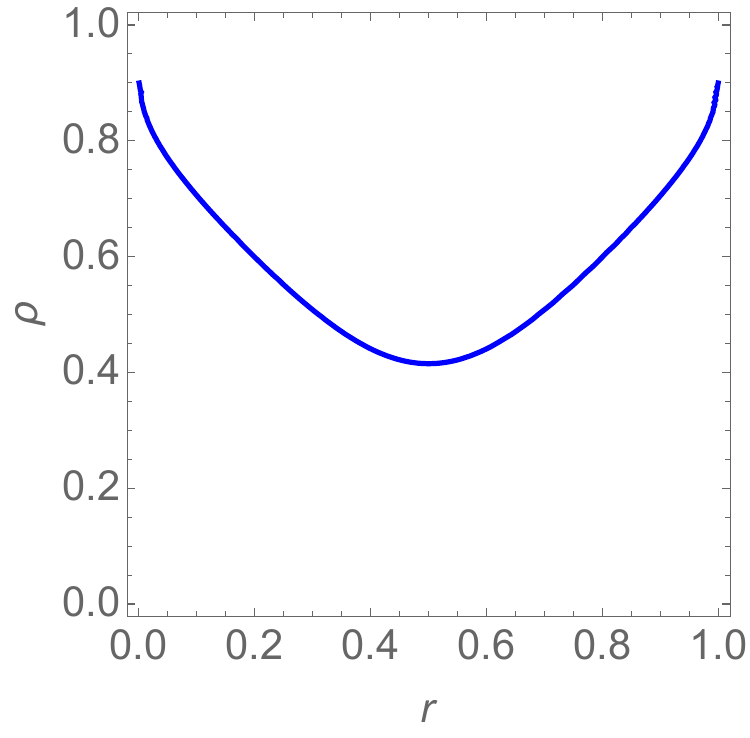}
\caption{The exponent $\rho$ as a function of the splitting parameter $r$ is implicitly determined by \eqref{rho}. The minimal value of the exponent given 
by \eqref{rho-min} is reached in the symmetric case of splitting into equal halves: $r=\frac{1}{2}$.}
\label{Fig:rr}
\end{figure}

Plugging the asymptotic \eqref{cAr} into \eqref{cr:small} we find that the exponent $\rho=\rho(r)$ is implicitly determined by relation (see also Fig.~\ref{Fig:rr})
\begin{equation}
\label{rho}
r^{\rho - 1} + (1-r)^{\rho - 1} = 3
\end{equation}
The exponent depends on the parameter $r$, but it does not depend on the fragmentation rate $\lambda$. The minimal value of the exponent 
\begin{equation}
\label{rho-min}
\rho_\text{min} = 2-\log_2 3 = 0.4150374992788\ldots
\end{equation}
is achieved in the symmetric case of splitting into equal halves: $r=\frac{1}{2}$. Near the minimum
\begin{equation}
\rho=\rho_\text{min} + \frac{\log_2 3(\log_2 3-1)}{\ln 4}\,(1-2 r)^2+\ldots
\end{equation}
In the limit of very unequal splitting, $R\ll 1$, the exponent slowly approaches to unity: 
\begin{equation}
1-\rho \simeq \frac{\ln 2}{\ln (1/R)}
\end{equation}

We now apply the Laplace transform to \eqref{AF:discrete} and arrive at a neat functional equation
\begin{equation}
\label{LT:r}
3 C(s) = \lambda^{-1}C^2(s)+C(rs)+C(s-rs)
\end{equation}
The small-mass tail \eqref{cAr} is reflected by the large $s$ asymptotic of the Laplace transform:
\begin{equation}
\label{LTr:large}
C(s) \simeq A(r)\,\frac{\Gamma(1-\rho)}{s^{1-\rho}}
\end{equation}

The Laplace transform $C(s)$ is expected to have the same type of singularity [see Eq.~\eqref{C:sing}] as in the model with random splitting: 
\begin{equation}
\label{Cr:sing}
C(s) -a \simeq -b\sqrt{s-s_*}
\end{equation}
as $s\downarrow s_*$. Plugging \eqref{Cr:sing} into \eqref{LT:r} and equating terms of the order $\sqrt{s-s_*}$ we obtain $a=3\lambda/2$, the same constant factor as in Eq.~\eqref{C:sing}. If the singular behavior is indeed given by \eqref{Cr:sing}, the large-mass behavior would read
\begin{equation}
\label{r:large}
c(x)\simeq \frac{b}{\sqrt{4\pi}}\,x^{-3/2}\,e^{s_* x}
\end{equation}
Performing the transformation \eqref{transform} one eliminates the dependence on $\lambda$; the dependence on the splitting parameter remains. One finds $b=B(r)\sqrt{4\pi \lambda}$ and $s_*=-\sigma(r)\lambda$, so Eq.~\eqref{r:large} acquires a more precise form
\begin{equation}
\label{r:large-better}
c(x)\simeq B(r)\sqrt{\lambda}\,x^{-3/2}\,e^{-\sigma(r) \lambda x}
\end{equation}

The transformation \eqref{transform} requires one to re-scale the Laplace transform
\begin{equation}
\label{Lap-F}
\begin{split}
& C(s)=\lambda\Phi(S), \quad s = \lambda S\\
& \Phi(S) = \int_0^\infty dX\, e^{-SX} F(X)
\end{split}
\end{equation}
Indeed, this rescaling recasts \eqref{LT:r} into equation 
\begin{equation}
\label{LT:Phi}
3  \Phi(S)  =  \Phi^2(S)+ \Phi(rS) + \Phi(S-rS)
\end{equation}
independent on $\lambda$. Recalling that $C(s_*)=3\lambda/2$ and $s_*=-\sigma(r)\lambda$ we obtain 
\begin{subequations}
\label{Phi}
\begin{equation}
\label{Phi-1}
\Phi[-\sigma(r)]=\frac{3}{2}
\end{equation}
Using \eqref{Phi-1} and specializing \eqref{LT:Phi} to $S_*=-\sigma(r)$ yields 
\begin{equation}
\label{Phi-2}
\frac{9}{4}  =  \Phi[-r\sigma(r)]+ \Phi[-(1-r)\sigma(r)]
\end{equation}
\end{subequations}
Neat relations \eqref{Phi} are not particularly useful since the scaled Laplace transform $\Phi(S)$ is unknown, and furthermore, non-universal (it depends on the parameter $r$). 

One may hope that the moments are more tractable analyticaly. The recurrence \eqref{Amn:r} leads to the functional equation
\begin{equation}
\label{Az:eq-r}
 \mathcal{A}(z)  =  \mathcal{A}^2(z) + \mathcal{A}(rz) + \mathcal{A}(z-rz)
\end{equation}
for the generating function \eqref{Az:def} of the moments. The asymptotic behavior \eqref{r:large-better} of the mass distribution implies the asymptotic behavior
\begin{equation}
\label{An:asymp-r}
A_n \simeq B(r)\,\Gamma\!\left(n-\tfrac{1}{2}\right)\,\sigma^{\frac{1}{2}-n}
\end{equation}
of the moments. Since $\Gamma\!\left(n-\tfrac{1}{2}\right)/\Gamma(n+1)\simeq n^{-3/2}$ when $n\gg 1$, we can re-write \eqref{An:asymp-r} in the form
\begin{equation}
\label{An:asymp-r2}
A_n \simeq \sqrt{\sigma(r)}\,B(r)\,\frac{n!}{n^{3/2}}\,[\sigma(r)]^{-n}
\end{equation}
resembling the asymptotic \eqref{An:asymp} of the moments in the model with random splitting. Therefore, the generating function \eqref{Az:def} converges when $0<z<\sigma(r)$ and exhibits the singular behavior
\begin{equation}
\label{Az:sing-r}
\mathcal{A}(z) = \frac{1}{2}-\sqrt{4\pi\sigma(r)}\,B(r)\,\sqrt{1-\frac{z}{\sigma(r)}}+\ldots
\end{equation}
when $z\uparrow \sigma(r)$. The leading square-root singularity is implied (cf. \cite{Flajolet}) by the asymptotic \eqref{An:asymp-r2}. The regular term, $\mathcal{A}[\sigma(r)] = \frac{1}{2}$, is derived by substituting the expansion \eqref{Az:sing-r} with undetermined $\mathcal{A}[\sigma(r)]$ into the functional equation \eqref{Az:eq-r} and equating the leading singular terms. 

The model with deterministic splitting is significantly more challenging than the model with random splitting. Indeed, the determination of the steady-state mass distribution requires solving functional equations \eqref{LT:Phi} and \eqref{Az:eq-r}. We established the qualitative behaviors of the steady-state mass distribution in the limits of small and large mass. The small-mass behavior, Eq.~\eqref{cAr}, is algebraic. We have computed the exponent \eqref{rho}, while the amplitude $A(r)$ remains unknown. In the large-mass asymptotic, Eq.~\eqref{r:large-better}, the overall amplitude $B(r)$ and the amplitude $\sigma(r)$ in the exponential term are unknown. 

Even linear functional equations are rarely solvable. The elegance of the nonlinear functional equations \eqref{LT:Phi} and \eqref{Az:eq-r} may inspire the hope of analytical advance. The symmetric case of splitting into equal halves, $r=\frac{1}{2}$, is the simplest. Equation \eqref{LT:Phi} becomes
\begin{equation}
\label{Phi2}
3\Phi(S)  =  \Phi^2(S)+ 2\Phi(S/2)
\end{equation}
The Laplace transform is determined for $S\geq - \sigma$. [We shortly write $ \sigma\equiv \sigma\big(\frac{1}{2}\big)$]. Using $\Phi(-\sigma)=\frac{3}{2}$, see \eqref{Phi}, we can compute $\Phi(-\sigma/2)$ from \eqref{Phi2}, and then $\Phi(-\sigma/4)$, $\Phi(-\sigma/8)$, etc.  Substituting $\Phi(-\sigma/2^n)=3\phi_n$ into Eq.~\eqref{Phi2} yields the nonlinear difference equation
\begin{equation}
\label{phi:eq}
\phi_{n+1} = \frac{3}{2}\, \phi_n (1-\phi_n), \qquad \phi_0 = \frac{1}{2}
\end{equation}

Equation \eqref{Az:eq-r} simplifies to
\begin{equation}
\label{Az2}
\mathcal{A}^2(z) = \mathcal{A}(z)  - 2\mathcal{A}(z/2)
\end{equation}
when $r=\frac{1}{2}$. Substituting $\mathcal{A}(\sigma/2^n)=a_n$ 
into Eq.~\eqref{Az2} we deduce the nonlinear difference equation
\begin{equation}
\label{a:eq}
a_{n+1} = \frac{1}{2}\, a_n (1-a_n), \qquad a_0 = \frac{1}{2}
\end{equation}

Equations \eqref{phi:eq} and \eqref{a:eq} are special cases of the logistic map, a very simple equation 
\begin{equation}
\label{logistic:eq}
x_{n+1} = \mu x_n (1-x_n)
\end{equation}
with very complicated behaviors arising when $\mu$ increases \cite{chaos}; e.g., chaos occurs when $\mu$ passes $3.56995\ldots$. The exact solutions of \eqref{logistic:eq} were found only in three special cases: $\mu=-2, 2, 4$, see \cite{logistic70,logistic10,logistic20}. In \eqref{phi:eq} and \eqref{a:eq}, we have $\mu=\frac{3}{2}$ and $\mu=\frac{1}{2}$.

\section{Homogeneous Random Splitting}
\label{sec:hom}

Here we consider a class of models with homogeneous random splitting---we assume that the total splitting rate of the cluster of mass $x$ is $\lambda x^\beta$. The master equation reads 
\begin{eqnarray}
\label{AF:hom}
\frac{\partial c(x,t)}{\partial t} &=& \int_0^x dy\,c(y,t)c(x-y,t) - 2N(t) c(x,t) \nonumber\\
                                             & + & 2\lambda \int_x^\infty dz\,\frac{z^\beta  c(z,t)}{z} - \lambda x^\beta c(x,t)
\end{eqnarray}

The governing equations describing the evolution in pure fragmentation models are linear and (formally) solvable for any breakup rate \cite{MZ85,MZ86}. For pure fragmentation models with homogeneous random splitting, the solutions behave very differently \cite{MZ85,MZ86,MZ87, Redner90,Ernst93} depending on whether $\beta\geq 0$ or $\beta<0$. A `shattering' transition in which mass is lost to dust composed of an infinite number of zero-mass particles occurs when $\beta<0$. One anticipates that aggregation-fragmentation processes described by Eqs.~\eqref{AF:hom} also behave very differently when $\beta\geq 0$ and $\beta<0$. Here, we analyze Eqs.~\eqref{AF:hom} with $\beta>0$. The models with $\beta<0$ are discussed Sec.~\ref{sec:minus}. 

We first argue that when $\beta\geq 0$, the mass distribution becomes stationary as $t\to\infty$. This feature allows us to focus on the analysis of the stationary mass distribution. 

It is natural to begin by studying the asymptotic behavior of the moments. Unfortunately, the moments do not generally satisfy closed equations when $\beta\ne 0$. For instance, the evolution equation for the cluster density
\begin{equation}
\label{Nb}
\frac{dN}{dt}=-N^2+\lambda M_\beta
\end{equation}
depends on the $\beta^\text{th}$ moment $M_\beta(t) = \int_0^\infty dx\,x^\beta c(x,t)$ of the mass distribution. The rate equation \eqref{Nb} is closed in just two cases: (i) $\beta=0$ when solution is given by \eqref{N:sol}, and (ii) $\beta=1$ when $\frac{dN}{dt}=-N^2+\lambda$ is also easily solvable and $N\to N(\infty)=\sqrt{\lambda}$.

One anticipates that the mass distribution becomes stationary when $\beta\geq 0$. The stationarity indeed occurs when $\beta=0$, as we argued in Sec.~\ref{sec:US}. For models with $\beta>0$, fragmentation of large clusters is more intense, so the flux to larger masses generated by pure aggregation is even more suppressed than for the model with $\beta=0$. 

Even for the model with $\beta=0$ studied in Sect.~\ref{sec:US}, more subtle behaviors may occur if the initial mass distribution has an algebraically decaying tail, $c(x,0)\sim x^{-a}$, with $2<a<3$. (The lower bound ensures that the mass density is finite, while the upper bound implies that the second moment is infinite: $M_2(0)=\infty$.) For compact initial mass distributions when all moments are finite, we have established explicit behaviors for a few moments that point to the stationarity. Below, we tacitly assume that the initial mass distribution is compact, or at least has a tail decaying faster than any power law. With this assumption, a reason for potentially unexpected behaviors is not the long tail in the initial mass distribution.

For the stationary mass distribution, Eqs.~\eqref{AF:hom} become 
\begin{eqnarray}
\label{mass:hom}
\big(2N+\lambda x^\beta\big) c = c*c+ 2\lambda \int_x^\infty dz\,\frac{z^\beta  c(z)}{z}
\end{eqnarray}
with $N = \sqrt{\lambda M_\beta}$ following from \eqref{Nb}. Recall that when $\beta=0$, the mass density diverges in the small-mass limit: $c\sim x^{-2/3}$, see Eq.~\eqref{small}. When $\beta>0$, the mass density remains finite in the small-mass limit. Letting $x=0$ in \eqref{mass:hom} we obtain 
\begin{equation}
\label{origin}
c(0) = \frac{\lambda}{N}\,M_{\beta-1} =  M_{\beta-1}\sqrt{\lambda/M_\beta}
\end{equation}
Equation \eqref{origin} is not an explicit formula since we do not have explicit expressions for the moments $M_{\beta-1}$ and $M_\beta$. 

The qualitative behavior of the stationary mass distribution changes when $\beta$ passes through $\beta_c=1$. When $0<\beta\leq 1$, the stationary mass distribution is maximal at $x=0$, and it is a decreasing function of $x$ on the entire mass range $0<x<\infty$. When $\beta>1$, the maximum occurs at a certain positive $x_*$. We do not prove the above assertions, but we provide strong evidence by analyzing the behavior of the mass distribution in the $x \to 0$ limit.

\subsection{$0<\beta<1$}

The convolution term scales as $x$, so it is sub-dominant (when $x\ll 1$) compared to the last term on the right-hand side of Eq.~\eqref{mass:hom}. A straightforward analysis gives 
\begin{equation}
\label{cx:less}
\frac{c(x)}{c(0)}= 1 - \frac{\lambda}{N}\left[\frac{1}{2}+\frac{1}{\beta}\right]x^\beta+\ldots
\end{equation}
when $x\ll 1$; the zero mass density is given by \eqref{origin}. The sub-sub-leading term omitted in \eqref{cx:less} depends on whether $\beta$ is smaller or larger than $\frac{1}{2}$. For instance, if $\frac{1}{2}<\beta<1$, a more accurate small-mass expansion  
\begin{equation}
\label{cx:less-sub}
\frac{c(x)}{c(0)}= 1 - \frac{\lambda}{N}\left[\frac{1}{2}+\frac{1}{\beta}\right]x^\beta+\frac{M_{\beta-1}}{2M_\beta}\,x+\ldots
\end{equation}
shows that the steady-state mass density is a decreasing function of $x$ when the mass $x$ is small; apparently, it is a decreasing function in the entire mass range $0<x<\infty$. 

\subsection{String model: $\beta=1$}
\label{subsec:1}

In a string model, the splitting rate is constant; the total splitting rate of a string of length $x$ is $\lambda x$. The string model is the most tractable model; its stationary mass distribution 
\begin{equation}
\label{c:exp}
c(x) = \lambda e^{-x\sqrt{\lambda}}
\end{equation}
was established in the first study of aggregation-fragmentation phenomena \cite{BT45}. The moments \eqref{Mn} read
\begin{equation}
\label{MA1}
M_n = \lambda^{-\frac{n-1}{2}}A_n\,, \qquad A_n = n!
\end{equation}

The convergence to the mass distribution \eqref{c:exp} has also been studied in considerable detail, see \cite{AB79}. 

\subsection{$1<\beta<2$}

The zero-mass density is still given by \eqref{origin}. The convolution term scales as $x$ and dominates the term $\lambda x^\beta c$ on the left-hand side of \eqref{mass:hom} in the $x\to 0$ limit. Taking this into account and performing the asymptotic analysis, we arrive at the small $x$ expansion identical to \eqref{cx:less-sub} where we should only exchange the sub-leading and the sub-sub-leading terms:
\begin{equation}
\label{cx:2more}
\frac{c(x)}{c(0)}= 1 + \frac{M_{\beta-1}}{2M_\beta}\,x -\frac{\lambda}{N}\left[\frac{1}{2}+\frac{1}{\beta}\right]x^\beta +\ldots
\end{equation}
The expansion \eqref{cx:2more} supports the above qualitative description that when $\beta>1$, the steady-state mass distribution $c(x)$ is a function with a single maximum.

\subsection{$\beta=2$}
\label{subsec:2}

The stationary mass distribution $c=c(x)$ satisfies
\begin{equation}
\label{mass2}
\big(2N+\lambda x^2\big) c = c*c+2\lambda \int_x^\infty dz\,z c(z)
\end{equation}
from which we deduce a non-linear ordinary differential equation for the Laplace transform \eqref{LT:def}:
\begin{equation}
\label{Lap:2}
\frac{d^2 C}{ds^2} - \frac{2}{s}\,\frac{d C}{ds} = \frac{2}{s}+
\frac{C^2-2NC}{\lambda}
\end{equation}
Equation \eqref{Lap:2} appears intractable, so let us look at the moments \eqref{Mn}. Using \eqref{mass2} we derive the recurrence 
\begin{equation}
\label{moments2}
 \lambda\,\frac{n-1}{n+1}\, M_{n+2} = \sum_{m=1}^{n-1} \binom{n}{m} M_mM_{n-m} 
\end{equation}
for $n\geq 2$. We also verify 
\begin{equation}
\label{M01}
N\equiv M_0 = \sqrt{\lambda M_2}, \qquad M_1 = 1
\end{equation}
There are no equations for $M_2$ and $M_3$. Specializing \eqref{moments2} to the smallest $n=2$ yields
\begin{equation}
\label{M4}
M_4 = 6/\lambda
\end{equation}
The moments are simple algebraic functions of $\lambda$:
\begin{equation}
\label{MA2}
M_n =  \lambda^{-\frac{n-1}{3}}A_n
\end{equation}
This ansatz is consistent with \eqref{moments2}--\eqref{M4} and it resembles expressions \eqref{MA} and  \eqref{MA1} for the moments in the models with $\beta=0$ and $\beta=1$.

The amplitudes satisfy the recurrence
\begin{equation}
\label{An:rec}
\frac{n-1}{n+1}\, A_{n+2} = \sum_{m=1}^{n-1} \binom{n}{m} A_m A_{n-m} 
\end{equation}
for $n\geq 2$. The `initial conditions' \eqref{M01} become
\begin{equation}
\label{A01}
A_0 = \sqrt{A_2}, \qquad A_1 = 1
\end{equation}
The initial conditions \eqref{A01} are insufficient: For arbitrary $A_2>0$ and $A_3>0$, one can determine $A_n = A_n(A_2,A_3)$ for $n\geq 4$ by 
recurrently solving \eqref{An:rec}:
\begin{equation}
\begin{split}
& A_4 = 6 \\
& A_5 = 12A_2 \\
& A_6 = 10\left[A_2^2+\tfrac{4}{3}A_3\right] \\
& A_7 = 30\left[3+A_2A_3\right] \\
& A_8 = 7\left[\tfrac{324}{5}A_2+4A_3^2\right], \quad \text{etc.}
\end{split}
\end{equation}
The large $n$ behavior of the moments 
\begin{equation}
\label{An-2}
A_n \simeq \frac{6(n+1)!}{B^{n+2}}
\end{equation}
involves yet undetermined amplitude $B$. The asymptotic \eqref{An-2} follows from \eqref{FX-large-2}; it can be also deduced by specializing the general asymptotic \eqref{An:beta} to $\beta=2$.  

The scaling behavior \eqref{moments2} of the moments is equivalent to the scaling behavior
\begin{equation}
\label{scaling2}
c(x) = \lambda^\frac{2}{3} F(X), \qquad X =  \lambda^\frac{1}{3} x
\end{equation}
of the mass distribution. Using \eqref{mass2} we deduce a governing equation for the scaled mass distribution 
\begin{equation}
\label{scaled-2}
\big(2A_0+X^2\big) F= F*F+ 2\int_X^\infty dZ\,Z F(Z)
\end{equation}
The small-mass expansion
\begin{subequations}
\begin{eqnarray}
\label{FX-small-2}
F(X) &=& \frac{1}{A_0}+  \frac{X }{2A_0^3}+\left(\frac{1}{4A_0^5} - \frac{1}{A_0^2}\right)\!X^2 \nonumber \\
&+& \left(\frac{5}{48A_0^7} - \frac{3}{4A_0^4}\right)\!X^3 + O(X^4) 
\end{eqnarray}
follows from \eqref{scaled-2}. Specializing the large-mass tail \eqref{FX} to $\beta=2$ gives
\begin{equation}
\label{FX-large-2}
F(X)\simeq 6 X e^{-BX}
\end{equation}
\end{subequations}
In contrast to the mass distribution for the string model, $F(X)=e^{-X}$, valid in the entire range, the mass distribution for the model with $\beta=2$ certainly differs from \eqref{FX-large-2} for small $X$ since $F(0)= \frac{1}{A_0}$.

\subsection{$\beta>0$: Scaling behavior}

The results of Sec.~\ref{subsec:1} and Sec.~\ref{subsec:2} suggest that the mass distribution exhibits scaling for arbitrary $\beta>0$:
\begin{equation}
\label{scaling-b}
c(x) = \lambda^\frac{2}{\beta+1} F(X), \qquad X =  \lambda^\frac{1}{\beta+1} x
\end{equation}
The scaled mass distribution obeys
\begin{equation}
\label{scaled-b}
\big(2A_0+X^\beta\big) F= F*F+ 2\int_X^\infty dZ\,Z^{\beta-1} F(Z)
\end{equation}
with $A_0 = \int_0^\infty dX\, F(X)$. The moments \eqref{Mn} read
\begin{equation}
\label{MAn}
M_n = \lambda^{-\frac{n-1}{\beta+1}} A_n, \qquad A_n = \int_0^\infty dX\, X^n F(X)
\end{equation}

A closed set of recurrence equations for amplitudes $A_n$ with integer $n$ can be written only when $\beta$ is a positive integer; generally, one ought to analyze an integro-differential equation \eqref{scaled-b}. If $\beta=k$ is a positive integer, the recurrence equations read 
\begin{equation}
\label{An-k:rec}
\frac{n-1}{n+1}\, A_{n+k} = \sum_{m=1}^{n-1} \binom{n}{m} A_m A_{n-m} 
\end{equation}
for $n\geq 2$. The problem is to fix $A_2,\ldots,A_{k+1}$. The first two amplitudes are
\begin{equation}
\label{A01-k}
A_0 = \sqrt{A_k}, \qquad A_1 = 1
\end{equation}

\subsection{Large-mass behavior}
\label{sub-large}

Let us look at the behavior of the mass distribution in the large-mass limit. We already know that if $\beta=0$, the tail is exponential with an algebraic pre-factor, Eq.~\eqref{large}; and if $\beta=1$, the tail, and the entire distribution, is pure exponential \eqref{c:exp}. It is natural to guess that the tail has a similar form
\begin{equation}
\label{CaB}
F(X)\simeq CX^{a-1} \exp(-B X)
\end{equation}
when $\beta\geq 0$. The exponent $a(\beta)$ and amplitudes $C(\beta)$ and $B(\beta)$ are independent on $\lambda$ since we consider re-scaled quantities. We now argue that
\begin{equation}
\label{aC}
a(\beta) = \beta, \qquad C(\beta) = \frac{\Gamma(2\beta)}{\Gamma(\beta)\,\Gamma(\beta)}
\end{equation}
for $\beta>0$. The prediction \eqref{aC} reduces to the known values $a(1) = 1$ and $C(1) = 1$ in the model with $\beta=1$. One cannot apply \eqref{aC} to the model with $\beta=0$ where $a(0) = \frac{1}{3}$ and $C(0) = \frac{9}{8\sqrt{\pi}}$. The model with $\beta=0$ is special: The steady-state mass distribution diverges in the small-mass limit; $c(0)$ is finite for models with $\beta>0$. 

To deduce \eqref{aC} we perform the Laplace transform of Eq.~\eqref{scaled-b} and find 
\begin{eqnarray}
\label{Phi-eq}
A_0 \Phi_0 = \frac{\Phi_0^2-\Phi_{\beta}}{2} + \frac{A_{\beta-1}-\Phi_{\beta-1}}{S}
\end{eqnarray}
where
\begin{equation}
\label{Phi-def}
 \Phi_\nu(S)=\int_0^\infty dY\,Y^\nu e^{-YS} F(Y), \quad A_\nu =  \Phi_\nu(0)
\end{equation}

The tail \eqref{CaB} of the scaled mass distribution implies the asymptotic behavior 
\begin{equation}
 \label{Phi-asymp}
\Phi_\nu(S)\simeq \frac{C\Gamma(a+\nu)}{(S+B)^{a+\nu}}
\end{equation}
in the $S+B\to +0$ limit. Substituting \eqref{Phi-asymp} into \eqref{Phi-eq} and equating the most diverging terms we arrive at \eqref{aC}.

Combining \eqref{CaB}  and \eqref{aC} we see that when $\beta>0$, the scaled mass density decays according to 
\begin{equation}
\label{FX}
F(X)\simeq \frac{\Gamma(2\beta)}{\Gamma(\beta)\,\Gamma(\beta)}\,X^{\beta-1}\,e^{-B(\beta) X}
\end{equation}
in the large $X$ limit, leading to the asymptotic 
\begin{equation}
\label{An:beta}
A_n\simeq \frac{\Gamma(2\beta)}{\Gamma(\beta)\,\Gamma(\beta)}\,\frac{\Gamma(n+\beta)}{[B(\beta)]^{n+\beta}}\qquad\text{as}\quad n\to\infty
\end{equation}
The amplitude $B(\beta)$ that appears in \eqref{FX} and \eqref{An:beta} is unknown, apart from the solvable string model where $B(1)=1$. Indeed, in the string model, $F(X)=e^{-X}$ is valid in the entire range, not only for $X\gg 1$. We do not know how to extract $B(\beta)$ from the asymptotic analysis.

\section{Models with $\beta<0$}
\label{sec:minus}

Analyzing Eqs.~\eqref{AF:hom} in the $\beta<0$ range is much more challenging than when $\beta\geq 0$ due to the following reasons:
\begin{enumerate}
\item The steady state is never reached.
\item The system undegoes a shattering transition: the mass is lost, which is interpreted as the emergence of dust, an infinite number of zero-mass fragments having overall positive mass. The loss of mass begins at $t=+0$, i.e., the shattering transition is instantaneous. 
\end{enumerate}
These two features of the solutions of Eqs.~\eqref{AF:hom} are related: Instantaneous shattering, which occurs when $\beta<0$, leads to the absence of a steady state. 

The precise description of the instantaneous shattering transition and the evolving cluster mass distribution for aggregation-fragmentation processes with $\beta<0$ is lacking. A qualitative justification of the above properties relies on pure fragmentation models governed by 
\begin{eqnarray}
\label{F:hom}
\frac{\partial c(x,t)}{\partial t} = 2\int_x^\infty dz\,\frac{z^\beta  c(z,t)}{z} -  x^\beta c(x,t)
\end{eqnarray}
with $\beta<0$. Equations \eqref{F:hom} are linear, and they admit an analytical treatment. Some (known) analytical results are presented in Appendix~\ref{ap:fragm}. These results show that the mass loss to dust indeed begins at $t=+0$. Loss of mass is known as the shattering transition \cite{Filippov}, and it begins at $t=+0$. Instantaneous shattering transitions occur in numerous models of pure fragmentation \cite{MZ86,MZ87,Redner90, Ernst93}. The shattering transition, i.e., the formation of dust, an infinite number of zero-mass particles containing a finite mass of the system, is dual to the gelation transition \cite{book,Flory,Leyvraz03} which describes the formation of an infinite gel in aggregation processes. 

The competition between aggregation and fragmentation usually leads to nonequilibrium steady states \cite{BT45,Melzak57,AB79,strings,Vigil,book}. In the processes described by Eqs.~\eqref{AF:hom} with $\beta<0$, however, aggregation has no time to counterbalance instantaneous shattering. Fragmentation remains explosive, and a steady state is never reached: Evolution never stops, and eventually everything turns into dust. 

We limit ourselves to heuristic arguments. Superficially, Eqs.~\eqref{AF:hom} appear ill-defined: the mass of the dust is finite, and hence the number of zero-mass fragments is infinite. However, $N(t)$ in Eqs.~\eqref{AF:hom} accounts for the density of clusters of {\em finite} mass. Interestingly,
\begin{equation}
\label{N:beta}
N(t)=
\begin{cases}
\infty             & -1\leq \beta<0\\
\text{finite}    & \beta<-1
\end{cases}
\end{equation}

In Appendix~\ref{ap:fragm}, we present a precise form of \eqref{N:beta} for pure fragmentation processes  governed by Eqs.~\eqref{F:hom}. For aggregation-fragmentation processes described by Eqs.~\eqref{AF:hom}, the density of clusters is expected to be smaller than the density given by \eqref{N:beta}, but still infinite if $-1\leq \beta<0$. Generally, the qualitatively different behaviors \eqref{N:beta} above and below $\beta=-1$ is caused by the small-mass asymptotic 
\begin{equation}
\label{cx:small}
c(x,t)\simeq B(t) x^{-2-\beta}
\end{equation}

In Appendix~\ref{ap:fragm}, we derive \eqref{cx:small} and determine the amplitude $B(t)$ for pure fragmentation processes. We now argue that \eqref{cx:small} also holds in aggregation-fragmentation processes. The same small-mass asymptotic \eqref{cx:small} immediately emerges due to fragmentation when $t\ll 1$, and it is not suppressed by aggregation. Importantly, \eqref{cx:small} remains valid throughout evolution; only the amplitude $B(t)$ is affected by aggregation. Indeed, let us consider the behavior of 
\begin{equation}
M(x,t)=\int_x^\infty dx\,x c(x,t)
\end{equation}
in the $x\to +0$ limit. Only fragmentation contributes to $\frac{dM(x,t)}{dt}$ in this limit. Using Eqs.~\eqref{AF:hom}, one finds
\begin{equation}
\label{lim}
\lim_{x\to +0} \frac{dM(x,t)}{dt} = - \lambda \lim_{x\to +0} \left[x^2\int_x^\infty dz\,\frac{z^\beta  c(z,t)}{z}\right]
\end{equation}
The left-hand side (LHS) converges to $\dot M = \frac{dM}{dt}$. The right-hand side (RHS) vanishes when $\beta\geq 0$, so the mass is conserved for these models. When $\beta<0$, the RHS of \eqref{lim} is negative if the small-mass behavior is given by \eqref{cx:small}, namely, it approaches to $-\lambda B(t)/2$, implying that
\begin{equation}
\label{cx:small-AF}
c(x,t)\simeq -\frac{2\dot M}{\lambda}\, x^{-2-\beta}
\end{equation}

The asymptotic \eqref{cx:small-AF} implies that \eqref{N:beta} is applicable. Therefore, Eqs.~\eqref{AF:hom} are mathematically well-defined when the parameter $\beta$ is sufficiently small: $\beta<-1$. When $-1\leq \beta<0$, the aggregation-fragmentation process is expected to remain well-defined on the physical grounds. Equations \eqref{AF:hom} have a divergent loss term due to aggregation, $-2N(t)c(x,t)$ with $N=\infty$, which should be compensated by the divergent gain term due to aggregation. The challenge is to recast Eqs.~\eqref{AF:hom} into a mathematically consistent form when $-1\leq \beta<0$. 

Another challenge is determining the cluster distribution $c(x,t)$ in the large time limit. In aggregation processes, solutions often approach a mass-conserving scaling form $c(x,t)=t^{-2s}\Phi(x t^{-s})$. It is often feasible \cite{Ernst85a} to determine the exponent $s$ that characterizes growth of the typical mass, $x_\text{typ}\sim t^s$, and establish qualitative behaviors of the scaled distribution $\Phi(\xi)$. In fragmentation, solutions also often approach \cite{Redner90} a mass-conserving scaling form $c(x,t)=t^{2\sigma}\Psi(x t^{\sigma})$. 

In evolving aggregation-fragmentation processes, one can imagine the emergence of two characteristic mass scales and even more complicated outcomes. Here, we analyze the simplest scenario: We postulate that the cluster distribution is characterized by a single scale. We thus assume that the solution acquires the scaling form
\begin{subequations}
\label{scaling}
\begin{equation}
\label{cxt:scaling}
c(x,t)=\frac{N^2}{M}\,\Phi(\xi)
\end{equation}
in the scaling limit
\begin{equation}
\label{xt:scaling}
x\to\infty, \quad t\to \infty, \quad \xi=\frac{Nx}{M}
\end{equation}
\end{subequations}
The definitions of $N$ and $M$ are consistent with \eqref{scaling} if the scaled distribution satisfies the integral relations 
\begin{equation}
\label{m-01}
\int_0^\infty d\xi\,\Phi(\xi) =1, \quad \int_0^\infty d\xi\,\xi\Phi(\xi) =1
\end{equation}
Substituting \eqref{scaling} into Eqs.~\eqref{AF:hom} and dividing by $N^2/M$ one gets a long equation with
\begin{subequations}
\begin{equation}
\text{LHS} = \left(2\frac{\dot N}{N}-\frac{\dot M}{M}\right)\Phi + \left(\frac{\dot N}{N}-\frac{\dot M}{M}\right)\xi\,\frac{d\Phi}{d\xi} 
\end{equation}
\begin{eqnarray}
\text{RHS} &=&\lambda\left(\frac{M}{N}\right)^\beta \left[2\int_\xi^\infty d\eta\,\eta^{\beta-1}\,\Phi(\eta)-\xi^\beta \Phi(\xi)\right]  \nonumber \\
&+&N[\Phi*\Phi(\xi)- 2\Phi(\xi)] 
\end{eqnarray}
\end{subequations}
The LHS and RHS must scale similarly with time. This requirement fixes the dependence of $N$ and $M$ on time 
\begin{equation}
\label{NM}
N\simeq \nu\, t^{-1}, \qquad M\simeq \mu\, t^{-1-1/\beta}
\end{equation}
and reduces the governing equation for the scaled cluster distribution to 
\begin{eqnarray}
\label{Phi:eq}
\beta^{-1}\xi\,\frac{d\Phi}{d\xi} &=& (1-\beta^{-1}-2\nu)\Phi + \nu \Phi*\Phi \nonumber \\
&+&\Lambda \left[2\int_\xi^\infty d\eta\,\eta^{\beta-1}\,\Phi(\eta)-\xi^\beta \Phi\right]  
\end{eqnarray}
where $\Lambda=\lambda(\mu/\nu)^\beta$. 

The scaling ansatz \eqref{scaling} and the decay laws \eqref{NM} are consistent with \eqref{cx:small-AF} when the scaled distribution has the small $\xi$ asymptotic 
\begin{equation}
\Phi(\xi) \simeq C\xi^{-2-\beta}, \qquad C=\frac{2(1+\beta^{-1})}{\Lambda}
\end{equation}

The above formulae are consistent only when $\beta<-1$. In this region, the solution conjecturally acquires the scaling form. Relying on the scaling ansatz we deduced that the density of clusters and the mass of finite clusters decay algebraically, Eq.~\eqref{NM}, with simple decay exponents. Solving an integro-differential equation \eqref{Phi:eq} appears impossible. The amplitudes $\nu$ and $\mu$ in Eq.~\eqref{NM} also remain undetermined.

\section{Discussion}
\label{sec:disc}

We explored the steady states in the model with mass-independent aggregation and fragmentation rates. The model with random and uniform splitting is solved in Sec.~\ref{sec:US}. We computed the Laplace transform, derived the extremal behaviors of the steady-state mass distribution, and computed all its moments. The models with deterministic splitting, $x \to [rx, (1-r)x]$ with fixed $r\in (0,1)$, are studied in Sect.~\ref{sec:DS}. The steady-state mass distribution is qualitatively the same as in the model with uniform splitting. To complete the analysis, one must solve nonlinear functional equations \eqref{LT:Phi} and \eqref{Az:eq-r}. This appears impossible even in the simplest symmetric case of splitting into equal halves, $r=\frac{1}{2}$. We established the qualitative behaviors of the steady-state mass distribution in the limits of small and large mass. The small-mass behavior \eqref{cAr} is characterized by the exponent given by Eq.~\eqref{rho}, and unknown amplitude $A(r)$. The large-mass behavior \eqref{r:large-better} resembles behavior in the model with random splitting. The amplitudes $B(r)$ and $\sigma(r)$ appearing in \eqref{r:large-better}  seem impossible to determine analytically. 

The models with mass-independent aggregation rates and homogeneous splitting rates that vary as $(\text{mass})^\beta$ are well-defined when $\beta\geq 0$. These models appeared in the mathematical literature (see \cite{BLL} and references therein); the convergence to a stationary mass distribution is not surprising. The steady states in the models with $\beta\geq 0$ are qualitatively understood, albeit not in full detail; e.g., we have not computed the amplitude $B(\beta)$ in the large-mass tail \eqref{FX}. 

The shattering transition in pure fragmentation models with $\beta < 0$ is instantaneous, and it is not suppressed by mass-independent aggregation. The density of clusters of finite mass diverges when $-1\leq \beta<0$. The loss term in Eqs.~\eqref{AF:hom} accounting for aggregation contains the divergent cluster density, so Eqs.~\eqref{AF:hom} appear ill-defined in this region. Physically, however, the loss is compensated by the gain. Recasting  Eqs.~\eqref{AF:hom} into a mathematically consistent form in the $-1\leq \beta<0$ region is an important challenge. Superficially, the situation resembles the zero-viscosity limit in turbulence; see, e.g.,  \cite{BD99,Eric,Khanin07,Anna14}. 

Equations \eqref{AF:hom} are mathematically well-defined when $\beta<-1$, and we analyzed them using scaling arguments. Employing the scaling ansatz, we found that the cluster density is inversely proportional to time, $N \sim t^{-1}$. The mass density also decays algebraically in the large time limit, $M\sim t^{-1-1/\beta}$. These asymptotic behaviors, together with the scaling ansatz \eqref{scaling}, lead to the algebraic decay of all moments
\begin{equation}
M_n(t) \simeq \nu \left(\frac{\mu}{\nu}\right)^n m_n\, t^{-1-n/\beta}
\end{equation}
We do not know the amplitudes $\nu$ and $\mu$, and the rescaled moments 
\begin{equation*}
m_n = \int_0^\infty d\xi\,\xi^n\Phi(\xi)
\end{equation*}
[The first two rescaled moments are $m_0=m_1=1$ by convention, \eqref{m-01}.]

The instantaneous shattering transition is a spectacular phenomenon arising in the continuous-mass framework. In the discrete-mass framework, the mass is always conserved. For pure fragmentation processes with $\beta<0$, there is a rapid increase in the density of the smallest fragments, the monomers \cite{MZ86}. This behavior mimics the loss of mass in the continuous-mass framework. The governing equations for aggregation-fragmentation processes in the discrete-mass framework are well-defined for all $\beta$. Investigating these equations when $-1\leq \beta<0$ may shed light on the regularization of  Eqs.~\eqref{AF:hom} and the behavior of solutions in this range of $\beta$. 

The shattering transition underlying many fragmentation processes \cite{Filippov,MZ86,MZ87,Redner90} is dual to the gelation transition \cite{book,Flory,Drake,Leyvraz03} which describes the formation of an infinite gel in aggregation processes. An instantaneous shattering transition may occur in the aggregation-fragmentation processes we explored, whereas gelation is impossible in our processes with mass-independent aggregation rates. Employing aggregation rates that may cause gelation would allow one to explore the competition between gelation and shattering. Aggregation processes with rates proportional to the product of masses of colliding clusters undergo the gelation transition \cite{Mcleod62a,McLeod}, so combining these rates with homogeneous fragmentation rates with $\beta<0$ may lead to the emergence of an infinite gel and the dust of zero-mass particles. Exploring the fate of such processes is an interesting avenue for future work. 

Phenomena resembling shattering to the zero-mass occur in other domains, e.g., in the evolution of the photon energy spectrum in plasmas through Compton scattering. The Kompaneets equation \cite{Komp} governing the evolution leads to photon loss \cite{Komp-Zeldovich,Komp-Lev,Komp-Vel,Komp-Lev2,Komp-Lev3}, despite the fact that photon numbers are nominally conserved in Compton scattering. This phenomenon is interpreted as the accumulation of photons at negligible values of energy, and it is analogous to the formation of a Bose-Einstein condensate.

\appendix
\section{Pure fragmentation with $\beta<0$}
\label{ap:fragm}

Here, we consider a one-parameter class of pure fragmentation processes governed by Eqs.~\eqref{F:hom}. These models are exactly solvable \cite{MZ86}. We outline a few exact results for processes with $\beta<0$ and derive \eqref{N:beta} from \eqref{F:hom} in a representative case of the monodisperse initial condition
\begin{equation}
\label{IC}
c(x,0)=\delta(x-1)
\end{equation}
This situation was analyzed in considerable detail \cite{MZ86} for the more general $c(x,0)=\delta(x-\ell)$ monodisperse initial condition, from which one can determine the evolution starting from an arbitrary initial condition due to the linearity of Eqs.~\eqref{F:hom}. 

\begin{figure}
\centering
\includegraphics[width=8.1cm]{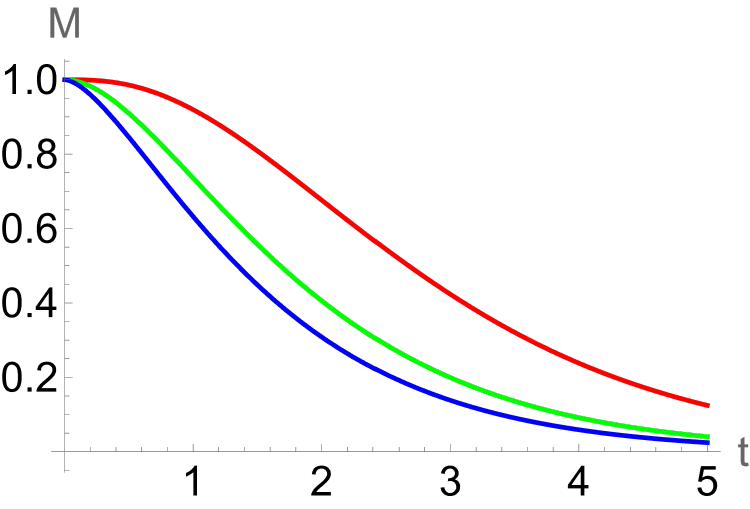}
\caption{The mass density (top to bottom) for $\beta=-1,-2, -3$. }
\label{Fig:M123}
\end{figure}

The solution of \eqref{F:hom} subject to \eqref{IC} has a singular part $e^{-t}\delta(x-1)$ describing intact clusters and a continuous part describing fragments with $x<1$. The continuous part is particularly simple when $\beta=-1$ and $\beta=-2$:
\begin{equation}
\label{Green-12}
c(x,t) = e^{-t}\times
\begin{cases}
2t-t^2 +t^2x^{-1} & \beta=-1\\
2t                       & \beta=-2
\end{cases}
\end{equation}
The mass loss begins at $t=+0$ as illustrated in Fig.~\ref{Fig:M123}. The explicit expressions for the mass density are particularly simple when $\beta=-1$ and $\beta=-2$:
\begin{equation}
\label{M12}
M(t) = e^{-t}\times
\begin{cases}
1+t+\frac{t^2}{2}  & \beta=-1\\
1+t                       & \beta=-2
\end{cases}
\end{equation}

The continuous part of the cluster density has been expressed \cite{MZ86} via the confluent hypergeometric function
\begin{equation}
\label{Green-beta}
c(x,t)=2t\,e^{-t}F\!\left[1+2 \beta^{-1};2;t-tx^\beta\right]
\end{equation}
in the general case. Using the $z\to -\infty$ asymptotic of the confluent hypergeometric function $F[a;b;z]$, one gets [cf. \eqref{cx:small}] the following small-mass behavior:
\begin{equation}
\label{cx:small-1}
c(x,t)\simeq \frac{2}{\Gamma\!\left(1-\frac{2}{\beta}\right)}\,t^{-\frac{2}{\beta}}e^{-t}x^{-2-\beta}
\end{equation}
The integral $\int_0^1 dx\,c(x,t)$ diverges in the small-mass limit when $-1\leq \beta<0$ and converges when $\beta<-1$, as asserted in \eqref{N:beta}. The asymptotic \eqref{cx:small-1} agrees with \eqref{cx:small};  we also find the amplitude $B(t)$ relevant for the evolution of the monodisperse initial condition \eqref{IC}. 

When $t\ll 1$, mass loss is the same for pure fragmentation and aggregation-fragmentation processes. The large time behaviors are drastically different. For instance, $M=(1+\lambda t)e^{-\lambda t}$ for the pure fragmentation process with $\beta=-2$. A much slower $M\sim t^{-1/2}$ decay occurs in the aggregation-fragmentation process with $\beta=-2$. For the same $\beta=-2$, the cluster density is $N=2\lambda t\,e^{-\lambda t}$ in pure fragmentation process, while for the aggregation-fragmentation process $N\sim t^{-1}$ when $t\gg 1$.

\bibliography{references-AF}

\end{document}